\documentclass[twocolumn]{aastex631}

\usepackage{amsmath}
\usepackage{amsfonts}
\usepackage{graphicx}
\usepackage{hyperref}
\usepackage{float}
\usepackage{multirow}

\usepackage{subfigure}

\begin{document}

\title{The hot circumgalactic medium in stacked X-rays: observations vs simulations}

\author{Skylar Grayson}
\affiliation{School of Earth and Space Exploration, Arizona State University, P.O. Box 876004, Tempe, AZ 85287, USA}

\author{Evan Scannapieco}
\affiliation{School of Earth and Space Exploration, Arizona State University, P.O. Box 876004, Tempe, AZ 85287, USA}

\author{Johan Comparat}
\affiliation{Max-Planck-Institut für extraterrestrische Physik (MPE), Gießenbachstraße 1, D-85748 Garching bei München, Germany}

\author{John A. ZuHone}
\affiliation{Center for Astrophysics,  Harvard and Smithsonian, 60 Garden St. Cambridge, MA 02138, USA}

\author{Yi Zhang}
\affiliation{Max-Planck-Institut für extraterrestrische Physik (MPE), Gießenbachstraße 1, D-85748 Garching bei München, Germany}

\author{Soumya Shreeram}
\affiliation{Max-Planck-Institut für extraterrestrische Physik (MPE), Gießenbachstraße 1, D-85748 Garching bei München, Germany}

\author{Marcus Brüggen}
\affiliation{Universität Hamburg, Hamburger Sternwarte, Gojenbergsweg 112, D-21029 Hamburg, Germany}

\author{Esra Bulbul}
\affiliation{Max-Planck-Institut für extraterrestrische Physik (MPE), Gießenbachstraße 1, D-85748 Garching bei München, Germany}

\begin{abstract}

Current cosmological simulations rely on active galactic nuclei (AGN) feedback to quench star formation and match observed stellar mass distributions, but models for AGN feedback are poorly constrained. The circumgalactic medium (CGM) provides a valuable laboratory to study this process, as its metallicity, temperature, and density distributions are directly shaped by AGN activity. Recent observations from the eROSITA instrument provide constraints on the CGM through measurements of extended soft X-ray emission. In this work, we generate synthetic eROSITA observations from the EAGLE and SIMBA cosmological simulations and compare them to observations of galaxies stacked by stellar mass, halo mass, and star-formation rate. SIMBA outperforms EAGLE in matching observed surface brightness profiles, but neither simulation achieves consistent agreement with observations across the full range of galaxy properties we studied. We find that variations in CGM X-ray emission between simulations are primarily driven by density differences at $R \lesssim 0.2 R_{200c} $, and temperature and metallicity changes at larger radii. These results highlight the need for further refinement of AGN feedback models in cosmological simulations and demonstrate the power of stacked X-ray observations as a tool for constraining feedback physics.
\end{abstract}


\section{Introduction}
\label{sec:intro}

The circumgalactic medium (CGM) is of vital importance to galaxy evolution. This diffuse and multiphase atmosphere around galaxies plays a central role in fueling star formation and is shaped by energetic feedback from stars and active galactic nuclei (AGN)  (see \cite{FaucherOh2023} for a recent, full review). AGN feedback, in particular, is thought to have a significant effect on the CGM in massive galaxies, as it drives outflows and injects enough energy to suppress gas accretion \citep{Scannapieco_2004, dimatteo2005, thacker2006, fabian2012, dubois2012, pillepich2018, Ma2022}. In cosmological simulations, AGN feedback is needed to be able to reproduce the observed populations of massive quiescent galaxies and is essential for galaxy quenching \citep[e.g.][]{dubois2014, pillepich2018, Simba, Schaye2015}. However, the details of AGN feedback and its impacts on the galactic environment are still largely unconstrained, and implementations of subgrid feedback processes in cosmological simulations vary significantly \citep{fabian2012, Oppenheimer_2021}. 

Studies of the hot phase ($T > 10^6$ K) of the CGM have become a useful tool for understanding the large-scale effects of AGN. This phase is governed primarily by gravitational and AGN heating, making it very sensitive to changes in AGN feedback prescriptions. Several signals can probe this hot phase, such as the thermal Sunyaev Zel'dovich (tSZ) effect (e.g. \cite{Spacek_2016, spacek2018, Grayson2023, hall_2019, Scannapieco_2008, meinke2021, meinke2023}) and X-ray absorption (e.g. \cite{bogdan2023, bhattacharyya2023, mathur2023}). X-ray emission also provides a promising pathway towards mapping out CGM properties (e.g. \cite{das2020, bertone2010, bogdan2013, Schellenberger2024, truong2020, Truong2023,zuhone2024}). 

With the recent data releases from the extended ROentgen Survey with an Imaging Telescope Array (eROSITA) \citep{brunner2022,merloni2024}, there has been a great interest in stacking X-ray emission to constrain the CGM around galaxies down to Milky Way masses. eROSITA was launched in 2019 on board the Spektrum-RoentgenGamma (SRG) orbital observatory and performed all-sky surveys with peak sensitivities in the soft X-ray \citep{merloni2012, Predehl2021}.  Initial work with the public Early Data Release eFEDS field data  \citep{brunner2022} stacked on optically selected galaxies, with  \citep{Chadayammuri2022} detecting CGM emission in $\approx$ 1600 galaxies with stellar masses 10.2 $<$ log($M_*/M_\odot$) $<$ 11.2, and \citep{Comparat_2022} stacking on $\approx$ 16,000 galaxies in the same field, detecting extended emission. More recently, the release of eRASS:4 \citep{merloni2024} has allowed for an analysis with a much larger sample size, with \cite{Zhang2024a} stacking $\approx$ 125,500 central galaxies, generating radial profiles, and detecting extended emission at high S/N. Follow-up work explored X-ray scaling relations in this sample \citep{Zhang2024b} and differences between star-forming and quiescent galaxies \citep{Zhang2024c}.

These results have also led to several studies comparing predictions from cosmological simulations to the observations. \cite{Chadayammuri2022} compared eROSITA observations to the cosmological IllustrisTNG \citep{springel2017} and EAGLE \citep{Schaye2015} simulations as analyzed in \cite{Oppenheimer_2020}. They found that simulations predict a stronger increase of X-ray luminosity with stellar mass, and that quenched galaxies in the simulations were dimmer than observations. However, the observational results used for these comparisons suffered from low sample sizes and potential star contamination, as explored in \cite{Zhang2024c}. \cite{Zhang2024b} compared the measured scaling relations against EAGLE, IllustrisTNG, and SIMBA \citep{Simba} simulations, but did not match galaxy property distributions or explore the impact of stacking by different parameters. Recently, \cite{Vladutescu2025} generated simulated surface brightness profiles using the Magneticum Pathfinder simulations\footnote{\url{http://www.magneticum.org/}}, finding generally poor agreement with observations from \cite{Comparat_2022} and \cite{Zhang2024a}. In \cite{lau2024} the CAMELS simulation suite \citep{Villaescusa2021} was compared to \cite{Zhang2024a} and \cite{Zhang2024b}, finding a preference for stronger AGN feedback, and no consistent alignment between simulations and observations. However, this work did not robustly match galaxy stellar mass distributions, and the simulations used had lower resolutions than fiducial versions leading to lower X-ray luminosities than those found in \cite{Zhang2024b}.

One of the more robust comparisons was carried out using TNG300, with \cite{shreeram2024} constructing a forward model that quantified uncertainties that arise from projection effects and misclassified centrals. They built on this by exploring the importance of galaxy selection when generating simulated profiles, with \cite{shreeram2025} finding that modifying the halo mass distribution in TNG300 stacks has a large effect on the stacked X-ray luminosity. The forward model developed in these works aligns well with observations from \cite{Zhang2024a}.  

As shown by the range of results presented here, there is no consensus regarding the success of simulations in replicating observed X-ray emission. The methods across these works differ, both in the pipelines used to generate simulated X-rays and in the sample selection process, some not robustly matching the observed galaxy sample properties. This variation in methodology makes it difficult to draw conclusions when comparing simulations, as many works only utilize one simulation or simulation suite. They also vary in their consideration of other X-ray contaminants which can play a large role in the detected emission, such as satellites \citep{shreeram2024}, AGN emission \citep{biffi2018}, and X-ray binaries \citep{Vladutescu2023}. 

In this work, we present the first robust comparisons between the eROSITA surface brightness profiles generated in \cite{Zhang2024a} and \cite{Zhang2024c} and the hydrodynamic simulations SIMBA \citep{Simba} and EAGLE \citep{Schaye2015}. These two simulations were chosen due to their very different treatments of AGN feedback, with SIMBA injecting energy primarily through kinetic jets and EAGLE utilizing thermal dumps. Despite the differences in feedback models, both simulations have been successfully calibrated to match global galaxy properties such as the stellar mass distribution and black-hole mass stellar mass relation \citep{Schaye2015, Simba}. There is thus a need to better understand how well each model does in matching other properties, such as the behavior of the diffuse baryons.  Here we use these simulations to model thermal X-ray emission from the CGM and we also model X-ray binary emission directly from star particles. By using the same pipeline for both simulations, we can understand the effect of the different feedback prescriptions evaluate the success of each in matching observations. We also stack galaxies by a range of properties and explore the underlying properties of the CGM gas, to understand what is driving differences in the observed X-ray surface brightness.  

The structure of this work is as follows. In Section \ref{sec2} we introduce the observational data and simulations used for this analysis. In Section \ref{sec3} we describe how our simulated eROSITA observations are generated and how we selected our simulated galaxy samples. In Section \ref{sec4} we show how CGM temperature, density, and metallicity vary between simulations, and explore how these variations are expected to contribute to X-ray surface brightness. We show radial surface brightness profiles stacked by both stellar and halo mass, and explore the contribution of X-ray binary (XRB) emission, the largest contaminant of X-ray observations of the hot CGM. In Section \ref{sec5} we explore the implications of these results in the context of comparisons to observations and future work constraining AGN feedback models.

\section{Data}\label{sec2}
\subsection{eROSITA}\label{sec2_1}
In this work, we compare simulated stacks to observational results from \cite{Zhang2024a} and \cite{Zhang2024c}. These observational results were derived from four all-sky surveys (eRASS:4) each lasting six months. The galaxies were selected from the SDSS DR7 spectroscopic catalog \citep{Strauss2002} with a stellar mass estimated from \cite{Chen2012}, SFR from \cite{Brinchmann2004}, and redshifts from \cite{Blanton2005}. The sample was split into four stellar mass bins ranging from $10^{10} M_\odot$ to $10^{11.25} M_\odot$ and five halo mass bins ranging from $10^{11.5} M_\odot$ to $10^{13.5} M_\odot$. The samples were further divided into star-forming and quiescent galaxies using the 4000 \r{A} break. The results we compare against here use only the galaxies identified to be central in their halos, labeled in \cite{Zhang2024a} and \cite{Zhang2024c} as the `CEN' sample. Table \ref{tab:observations} shows the counts of this sample, organized by stellar and halo mass bin, as well as their redshift ranges.

\begin{table}[h]
\caption{A summary of the observational galaxy samples we compare against. The samples are further divided into star-forming (SF) and quiescent (QU) galaxies. Adapted from Table 1 in \cite{Zhang2024c}.} 
    \label{tab:observations}
    \centering
    \begin{ruledtabular}
    \begin{tabular}{cccccc}
        log10($M_*/M_\odot$) & & redshift & & N &  \\
        min & max & min & max & N$_{SF}$ & N$_{QU}$\\
        \hline
        10 & 10.5 & 0.01 & 0.06 & 5226 & 2516\\
        10.5 & 11 & 0.02 & 0.1 & 13668 & 16485 \\
        11 & 11.25 & 0.02 & 0.15 & 8070 & 17466 \\
        11.25 & 11.5 & 0.03 & 0.19 & 4279 & 15663 \\
        \hline
        \hline
        log10($M_{200}/M_\odot$) & & redshift & & N & \\
        min & max & min & max & N$_{SF}$ & N$_{QU}$\\
        11.5 & 12 & 0.01 & 0.08 & 20206 & 5162 \\
        12 & 12.5 & 0.02 & 0.13 & 29791 & 11510 \\
        12.5 & 13 & 0.02 & 0.16 & 12103 & 18637 \\
        13 & 13.5 & 0.03 & 0.2 & 1788 & 18101 \\
        13.5 & 14 & 0.03 & 0.2 & - & 8209\\
        \hline
        
    \end{tabular}
    \end{ruledtabular}
    
\end{table}

\subsection{SIMBA}\label{sec2_2}

The first set of galaxy formation simulations we compared to observations was carried out using the SIMBA code. This code is built on the meshless finite-volume solver GIZMO, but differentiates itself in multiple ways, including its treatment of AGN feedback \citep{hopkins2014, Simba}, which is modeled in two distinct modes.
The first, associated with high Eddington accretion rates ($f_{\text{Edd}} = \dot{M}_{\text{BH}}/\dot{M}_{\text{Edd}}>0.2$), is described as a wind mode, with velocities given by 
\begin{equation}
v_{w} = 500 + 500 (\text{log} M_{\text{BH}}-6)/3 
\; \text{km s}^{-1},
\end{equation}
where $M_{\text{BH}}$ is the black hole mass. The winds are ejected at the interstellar medium (ISM) temperature with a mass loading given by 
\begin{equation} \label{bh_momentum}
\dot{P}_{\rm out} = 20 c \eta  \dot{M}_{\text{BH}},
\end{equation}
where $\eta = 0.1$ and $c$ is the speed of light. 

At low Eddington rates ($f_{\text{Edd}} \leq 0.2$) the AGN feedback takes on a jet mode, with ejection velocities of 
\begin{equation}
v_j = v_w + 7000 \; \text{log}(0.2/f_{\text{Edd}})\text{km s}^{-1}.
\end{equation}
The jet mode can only be triggered if the mass of the black hole $M_{\text{BH}}>10^{7.5} M_\odot$. The mass loading is set by the same momentum rates given in Equation \ref{bh_momentum}, and the gas temperatures are set to the virial temperature of the halo. For both the jet and wind modes, particles are ejected in a bipolar fashion perpendicular to the disk used to compute the black hole accretion. In association with the jet mode, SIMBA also includes additional feedback energy associated with X-rays produced in the accretion disk. This X-ray mode only occurs if $M_{\text{gas}}/M_* < 0.2$ and the host galaxy stellar mass is $ >10^9 M_\odot$. The amount of energy released is determined by equations 11 and 12 in \cite{Choi2012}, with half of the energy applied to radial outward momentum and the remainder added as heat \citep{Simba}.

SIMBA also models photoionization heating and radiative cooling with the GRACKLE 3.1 library, which includes metal cooling, non-equilibrium evolution, and self-shielding \citep{Smith2017}.

For our comparisons, we used the publicly-available SIMBA runs, choosing a box size of 50 Mpc h$^{-1}$ with 1024$^3$ gas and dark matter particles. This box size was run not only with all the above AGN feedback modes on (hereafter referred to as `SIMBA'), 
as well as a run with no AGN feedback (hereafter `SIMBA-NoAGN'). This code assumes a cosmology wherein $H=68$ km s$^{-1}$ Mpc$^{-1}$ and the matter, baryon, and vacuum densities are $\Omega_m=0.3$, $\Omega_b = 0.048$, and $\Omega_\Lambda= 0.7$ respectively \citep{Simba, Planck2016}. 

Galaxy and halo catalogs were generated using CAESAR, a yt-based package that uses a 6D friends-of-friends algorithm with a linking length 0.0056 times the mean interparticle spacing to identify structures \citep{Simba}. 

\begin{table*}[t]
\caption{A summary of the runs used in this work. All models are publicly available, and more detail on the runs can be found in \cite{Simba} (SIMBA), and \cite{Schaye2015} (EAGLE). }
    \label{tab:runs}
    \centering
    \begin{ruledtabular}
    \begin{tabular}{ccccccc}
        Label  & Box Size & Initial Gas/DM Particle Count & Cosmology & Feedback Treatment  \\
        \hline

        EAGLE & 50 Mpc cm & $752^3$ & Planck 2013 & EAGLE Fiducial, $\Delta T_{\text{AGN}} = 10^{8.5}$ \\
        EAGLE-AGNdT9  & 50 Mpc cm & $752^3$ & Planck 2013 & EAGLE Fiducial, $\Delta T_{\text{AGN}} = 10^9$ \\
        EAGLE-NoAGN  & 50 Mpc cm & $752^3$ & Planck 2013 & None\\
        SIMBA & 50 Mpc h$^-1$ & $1024^3$ & Planck 2015 & SIMBA Fiducial \\
        SIMBA-NoAGN & 50 Mpc h$^-1$ & $1024^3$ & Planck 2015 & None  \\
        \hline
    \end{tabular}
    \end{ruledtabular}
    
\end{table*}

\subsection{EAGLE}\label{sec2_3}

The other simulation we used for our comparisons is EAGLE \citep{Schaye2015}. The EAGLE simulations are based on the SPH and N-body code GADGET3 \citep{springel2005}, with an AGN feedback model that differs significantly from SIMBA. There is only one mode, and the feedback energy is injected thermally, as opposed to SIMBA's kinetic implementation. The energy injection rate is given by 
\begin{equation} \label{eagle_energy}
\dot{E}_{\text{BH}} = \epsilon_f \epsilon_r \frac{\dot{M}_{\text{BH}}}{1-\epsilon_r}c^2,
\end{equation}
where $\epsilon_f=0.15$ is the fraction of the radiated energy that coupled to the ISM and $\epsilon_r= 0.1$ is the radiative efficiency of the accretion disk \citep{Schaye2015}. The energy is deposited into the environment based on a reservoir model, where each black hole has a given energy in the reservoir ($E_{\text{BH}}$) which is increased each timestep following Eq.\ (\ref{eagle_energy}). The energy is released when it becomes large enough to heat each SPH neighbor by a given $\Delta T_{\text{AGN}}$.

The value of $\Delta T_{AGN}$ depends on the size and resolution of the box. For this work, we use the model Ref-L050N0752 (hereafter EAGLE), which consists of a box with a side length of 50 comoving Mpc, and 752$^3$ DM particles, and a value of $\Delta T_{\text{AGN}}=10^{8.5}$ K. We also use the model AGNdT9-L050N0752 (hereafter EAGLE-AGNdT9) which boosts $\Delta T_{AGN}=10^{9}$ K \citep{Schaye2015}. The AGNdT9 run also modifies the accretion parameterization such that significant AGN activity happens at higher black hole masses. While both runs match observed galaxy stellar mass functions, EAGLE-AGNdT9 predicts lower gas fractions on the group scale ($M_{500, \text{hse}}<$ $10^{14} M_\odot$), matching observations somewhat better than EAGLE. EAGLE also slightly overpredicts cluster X-ray luminosities, with EAGLE-AGNdT9 predicting lower luminosities more in line with observations,  although it is worth noting there are very few clusters in these boxes \citep{Schaye2015}. Thus the use of both runs allows us to explore the sensitivities of X-ray CGM emission to the AGN feedback strength within a given model. Finally, we use the NoAGN-L050N0752 model (hereafter EAGLE-NoAGN), which turns off the AGN feedback. 

Cooling and radiative heating in these simulations are modeled element by element for H, He, C, N, O, Ne, Mg, Si, S, Ca, and Fe, based on rates from CLOUDY version 07.02 \citep{ferland1998, wiersma2009}. All three EAGLE runs used here assume a cosmology wherein $H=67.77$ km s$^{-1}$ Mpc$^{-1}$, the matter, baryon, and vacuum densities are $\Omega_m=0.307$, $\Omega_b = 0.04825$, and $\Omega_\Lambda= 0.693$ respectively \citep{Schaye2015, Planck2014}.

We identified EAGLE galaxies and halos using the publicly released catalogs \citep{McAlpine_2016}. In this work, we use the Subhalo catalog, consisting of subhalos identified using the SUBFIND algorithm \citep{springel2001, dolag2009}. Particles are not shared between subhalos, and there can be multiple subhalos within a given halo. The central galaxy is then identified as the subhalo containing the particle with the lowest gravitational potential in the halo. From this identification, one can determine galaxy properties, although we note that this method of identifying particles associated with galaxies will include diffuse particles at large distances, which would contribute observationally to intra-cluster/intra-group light, and thus are often excluded by photometric methods \citep{Schaye2015}. We also chose the Subhalo catalog over others as it defines galaxies in a similar way to the friends-of-friends algorithm in CAESAR, allowing for a more robust comparison between the SIMBA and EAGLE results.

\begin{figure}[ht!]

        \centering
        \includegraphics[width=0.5\textwidth]{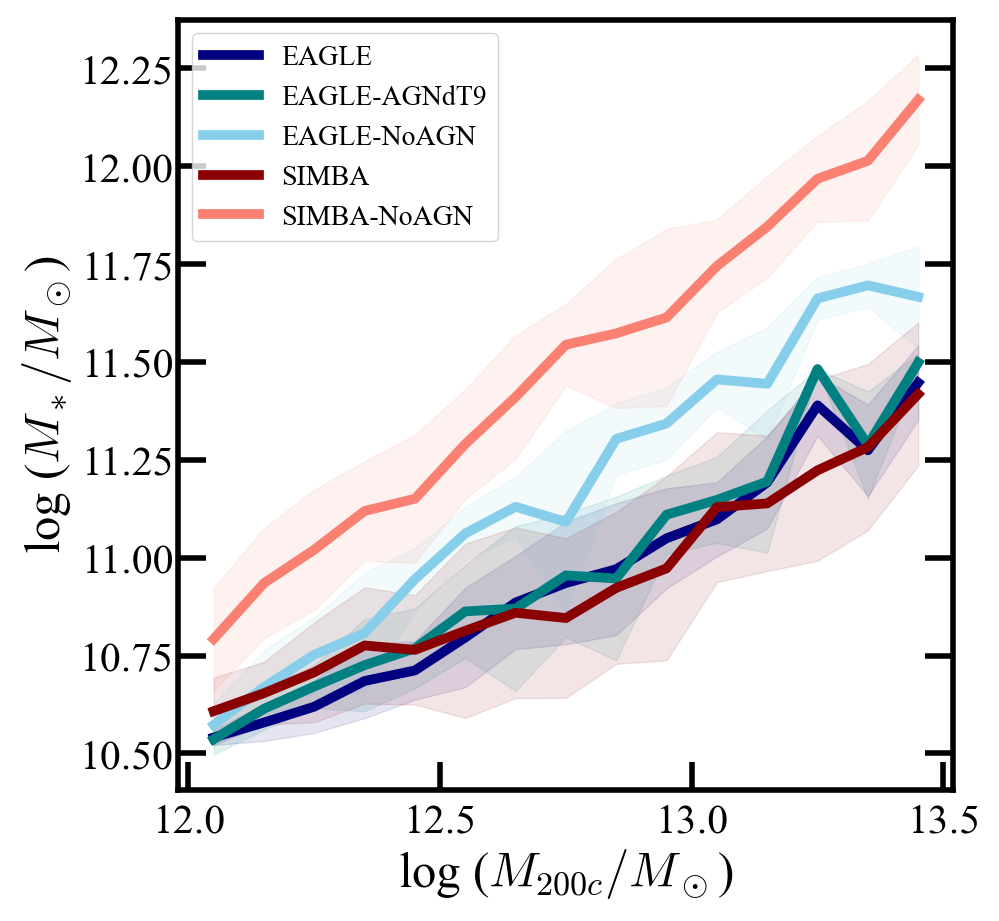}

    \caption{Halo-stellar mass relationship for the five simulations used in this work. Here we show the average central stellar mass for each 0.1 dex halo mass bin. Shaded regions represent the one sigma spread in each halo mass bin.}
    \label{fig:stell-halo relation}
\end{figure}

Table \ref{tab:runs} summarizes the runs used for this analysis, and Figure \ref{fig:stell-halo relation} shows their halo-stellar mass relationships. Note that, because massive galaxies cannot be quenched without AGN feedback, the two -NoAGN runs do not represent realistic models of the universe, as they do not reproduce galaxy stellar mass functions or stellar-halo mass relationships.  However, we still include them in our analysis as a) we make halo as well as stellar mass cuts in our galaxy selection (see Section \ref{gal_selection}), and b) the comparison allows us to better understand exactly how AGN feedback is impacting the CGM, as explored more in Section \ref{sec4-cgm}.

\begin{figure}[ht!]

        \centering
        \includegraphics[width=0.5\textwidth]{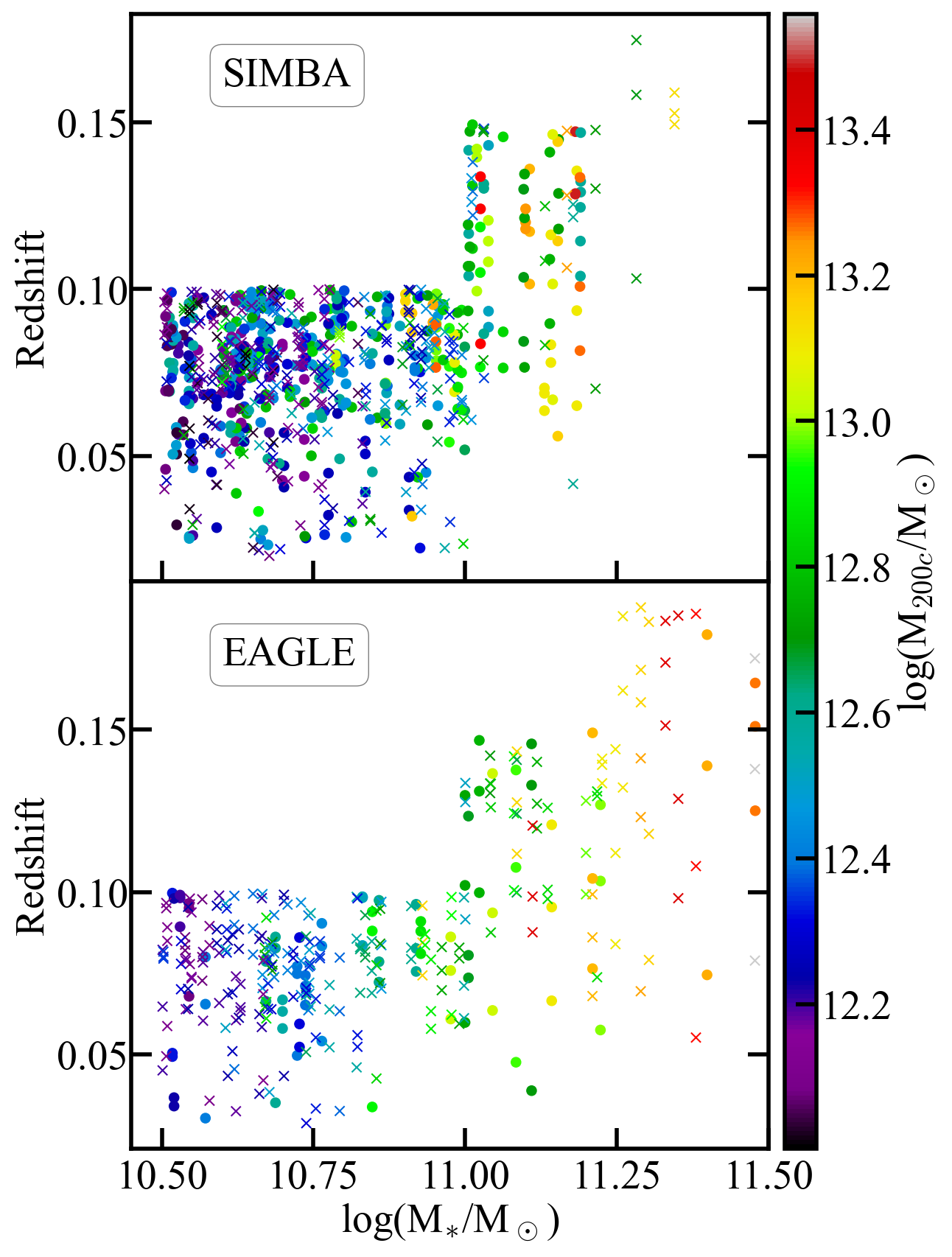}

    \caption{Galaxy sample characteristics for the SIMBA (top) and EAGLE (bottom) analysis prior to resampling. Here we see redshift against stellar mass, with galaxies color-coded by halo mass. Star-forming galaxies (those with a sSFR $>$0.01) are represented with an x, while quiescent galaxies are shown as circles. Appendix \ref{app:counts} gives the sample sizes for the different simulated galaxy samples.}
    \label{fig:galaxy_sample_characteristics}
\end{figure}

\section{Analysis} \label{sec3}

\subsection{Simulated Galaxy Selection}\label{gal_selection}

To match the observed galaxy sample presented in Table \ref{tab:observations}, we created simulated observations of each  $z=0.1$ snapshot along the $x$, $y$, and $z$ directions. For each galaxy in the simulation box, we assigned a redshift for each synthetic observation by randomly choosing from the observed redshift distribution in \cite{Zhang2024c} based on the galaxy stellar mass bin it falls in (with bin widths of 0.5 dex). We selected only galaxies identified by the halo finder as being central, so as to compare against the CEN sample from \cite{Zhang2024a, Zhang2024c}, which excludes satellites from the stacked results. The resulting redshift distribution is shown in Figure  \ref{fig:galaxy_sample_characteristics}, ranging from z $\approx$ 0.02 to 0.2. We elect to use only the z=0.1 snapshot for this analysis as the purpose of assigning different redshifts is purely to account for PSF effects that impact the observed surface brightness profiles, and we expect very little evolution for the galaxies in this redshift range.

During the stacking process, the galaxies were resampled to account for observational biases that impact the shape of the galaxy stellar mass function. Galaxies were sorted into 0.1 dex bins, and the simulated galaxies were resampled to create a new sample with $N = 0.2 N_{obs}$, following the same stellar mass distribution of the observational sample. This resampling process ensured consistency with both the mean and the spread of the observations. The new simulated galaxy sample was then used to generate stacks. 

Recent work has shown the importance of halo mass distribution on stacked X-ray results \citep{shreeram2025}. As there are large uncertainties involved in determining the halo masses of galaxies in the observational sample, we have elected not to attempt to match the observed halo mass distribution, although we do exclude galaxies with M$_{200c}$ $>10^{14}$ or $<10^{12}$ from our analysis. Figure \ref{fig:stell-halo relation} shows the halo mass-stellar mass relationship for the five simulations used in this work. Without AGN feedback, this relationship is changed significantly, with halos hosting much larger galaxies as star formation is unimpeded by AGN activity. Section \ref{sec4_2} will explore halo mass selections more in-depth.

In \cite{Zhang2024c}, galaxies are divided into star-forming and quiescent using measurements of the 4000 \r{A} break \citep{Brinchmann2004}. Matching this cut in simulations is difficult, and so we elect instead to divide the galaxies based on their sSFR such that galaxies with sSFR $>$ 0.01 Gyr$^{-1}$ are labeled as star-forming, and galaxies with sSFR $<$ 0.01 Gyr$^{-1}$ are labeled as quiescent. Figure \ref{fig:galaxy_sample_characteristics} shows the properties of the galaxies. Full galaxy counts for these subsamples can be found in Appendix \ref{app:counts}. Due to the box size, the number of galaxies is very small, particularly at the high mass end (i.e. both EAGLE and SIMBA have only 6 quiescent galaxies in the 11.25 $<$ log($M_* / M_\odot$) $<$ 11.5 bin. For this reason, we focus our analysis on the combined star-forming and quiescent sample. Appendix \ref{app:xrb_profiles} shows the results for the separated samples, as a preliminary point of comparison for studies concerned with star formation activity.

\subsection{X-ray Forward modeling}\label{xraymodel}

To produce synthetic X-ray images of our simulated galaxies, we use pyXSIM \citep{zuhone2016} to generate lists of photons produced by hot plasma within 2000 kpc of each galaxy. pyXSIM generates photons under collisional ionization equilibrium via Monte-Carlo sampling of the Astrophysical Plasma Emission Code (APEC) \citep{Smith2001}. The resulting X-ray emission is thus dependent on the temperature, metallicity, and density of each gas particle. We exclude any particles with $n_H > 0.1$ cm$^{-3}$ as those are ISM particles artificially pressurized to prevent runaway star formation in these relatively low-resolution simulations (see \cite{springel2003}). These are generally very cool particles that would not contribute to X-ray emission regardless. Once we have the photon lists around each galaxy, we use pyXSIM to create event lists, or line-of-sight projections including Doppler shifting due to velocities of the fluid elements \citep{zuhone2016}. For all analyses, we adopt the cosmology from \cite{planck2020}, which sets $H_0 = 67.74$ km/s/Mpc and $\Omega_M = 0.3089$. Note that cosmic ray (CR) feedback does not play a role in any of the simulations considered here \citep[as opposed to recent work e.g.][]{Butsky2018,buck2020,thomas2022}.  Thus we do not include additional X-ray emission from inverse compton scattering of cosmic microwave background photons by CR electrons as in \citep{Hopkins25, Lu25}.

We used the Simulated Observations of X-ray Sources (SOXS)\footnote{\url{https://hea-www.cfa.harvard.edu/soxs/}} package to create simulated eROSITA observations. The SOXS instrument simulator had not been used for eROSITA prior to our analysis, so we built in the functionality using instrument files (PSF, ARF, RMF, backgrounds) from the SIXTE analysis tool \citep{Dauser2019}.

\begin{figure*}[ht!]

        \centering
        \includegraphics[width=0.96\textwidth]{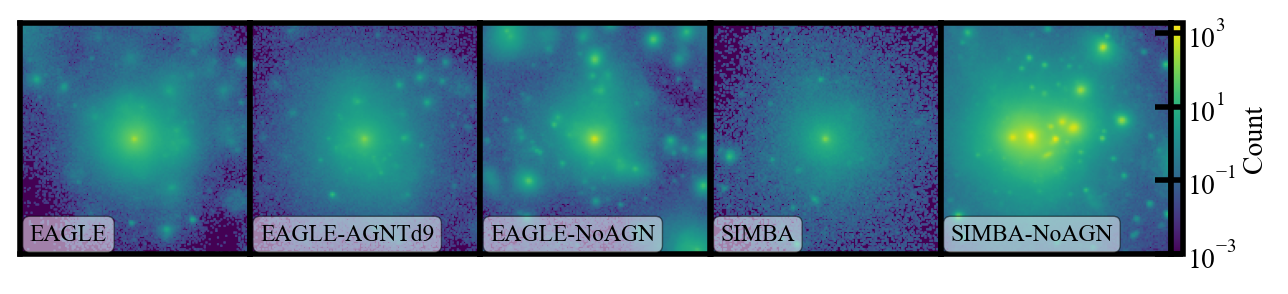}

    \caption{Stacked synthetic observations of hot CGM emission for galaxies with $11< \;$log$(M_*/M_\odot)<11.25$ for (left to right) EAGLE, EAGLE-AGNdT9, EAGLE-NoAGN, SIMBA, SIMBA-NoAGN. Each box is 22.5 arcminutes across ($\approx$ 2.5 Mpc at $z=0.1$).  }
    \label{fig:hotcgmmaps}
\end{figure*}

SOXS generates images from pyXSIM event files in a multi-step process. First, it uses the auxiliary response files to determine which events will be detected. The detected events then undergo a projection onto the detector plane with PSF blurring as determined by the PSF model supplied by SIXTE. SOXS then provides the option to add instrumental and astrophysical backgrounds and foregrounds, which we discuss below. Finally, the event energies are convolved according to the instrument responses. eROSITA consists of seven independent mirror assemblies, so we model each mirror assembly as a separate instrument in SOXS and stack the results. As we are comparing against fully cleaned and background subtracted observations, we do not include any foreground or background models, instead generating emission purely from the gas within 2,000 kpc of each galaxy. We observe each galaxy for 1,000ks, which is significantly longer than the average exposure of $\approx$ 550 seconds from eRASS:4. This was done to reduce noise given the low sample sizes. Figure \ref{fig:hotcgmmaps} shows stacked counts for galaxies in each simulation box with $11< \;$log$(M_*/M_\odot)<11.25$.

We determine the surface brightness following the procedure initially outlined in \cite{Comparat_2022} and used in \cite{Zhang2024c}. We calculate the weight $w_i$ of each event with energy $E_{\rm obs}$ (in keV) using

\begin{equation}
w_i = \frac{1.602177 \times 10^{-9} E_{\rm obs} \times (1+z)}{ARF(E_{\rm obs})}\frac{4 \pi d_L^2}{t_{\rm exp}},
\end{equation}
where $z$ is the redshift of our simulated observation, $ARF$ represents the area response function for all seven mirror assemblies, $d_L$ is the luminosity distance corresponding to redshift $z$, $t_{\rm exp}$ is the exposure time, and the constant out front converts to the correct energy units. From this, we can separate the events into radial shells with the widths set in \cite{Zhang2024c}\footnote{The radial bin boundaries are: 0, 10, 30, 50, 75, 100, 200, 300,
400, 500, 600, 700, 800, 900, 1000, 1100, 1200, 1300, 1400, 1600,
1800, 2000 kpc.} and calculate a mean surface brightness for the galaxy stack as
\begin{equation}
S_X = \Sigma \frac{w_i}{A_{\text{shell}}N_g},
\end{equation}
where $A_{\text{shell}}$ is the area of the shell and $N_g$ is the number of galaxies in the stack. 
As we generated X-rays directly from thermal emission in the CGM without any backgrounds or foregrounds, we perform no background subtraction and focus our analysis to within $\approx$ 1000 kpc.

\begin{figure}[ht!]

        \centering
        \includegraphics[width=0.4\textwidth]{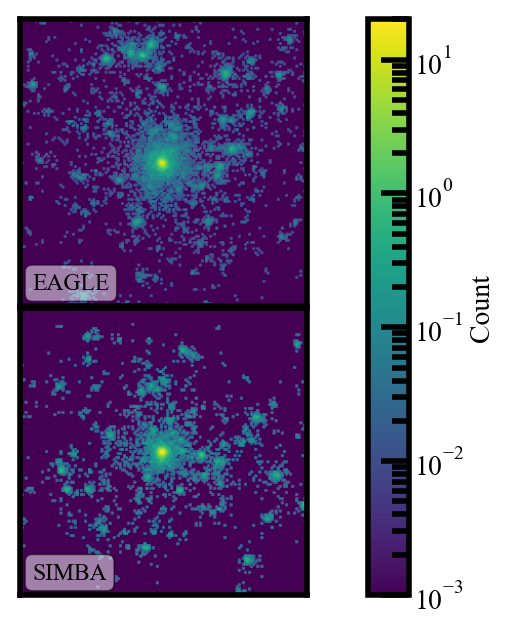}

    \caption{Stacked synthetic observations of XRB emission for galaxies with $11< \;$log$(M_*/M_\odot)<11.25$ in EAGLE (top), and SIMBA (bottom). Each box is 22.5 arcminutes across ($\approx$ 2.5 Mpc at $z=0.1$).  }
    \label{fig:xrbmaps}
\end{figure}

\begin{figure}[ht!]
\includegraphics[width = 0.45\textwidth]{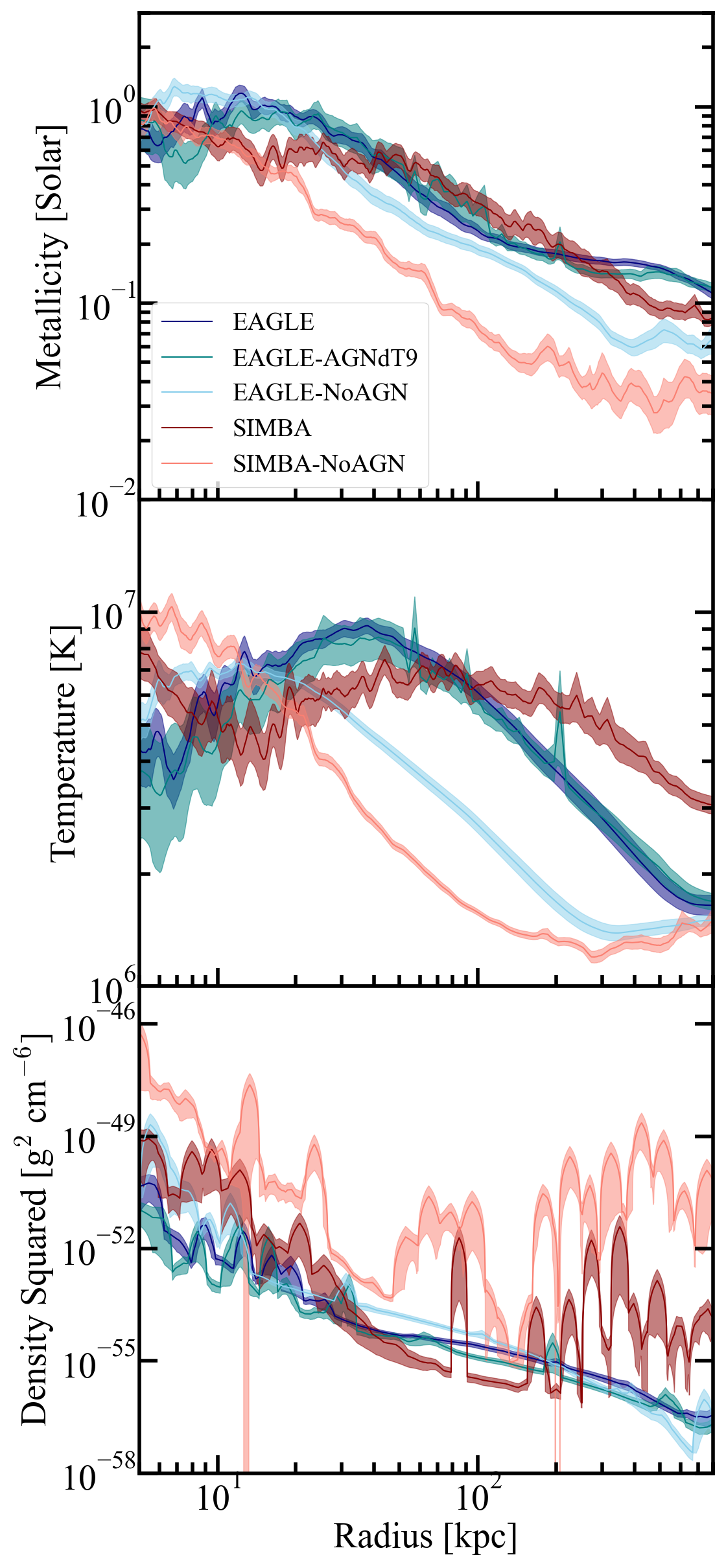}
    \caption{Radial profiles of the average mass-weighted metallicity (top), temperature (center), and density squared (bottom) of M31-type central galaxies ($11<\text{log}(M_*/M_\odot)<11.25$, $12<\text{log}(M_{200c}/M_\odot)<14$). These results only include gas particles with T$>10^6$ K, as cooler particles will not significantly contribute to the soft X-ray emission (see Figure \ref{fig:apec}). Error bars represent the 95$\%$ confidence interval generated by bootstrapping the radial profiles for all galaxies in the sample. Metallicities are given as the mass fraction of elements heavier than helium normalized to a solar value of 0.02.}
    \label{fig:gas_properties}
\end{figure}

\subsection{XRBs}\label{sec3_1}

The final element included in our analysis is emission from X-ray Binaries (XRBs). XRBs are the largest contaminant of X-ray observations of the hot CGM, so we would like to understand the relative contributions of XRB emission in our simulations. The XRB population is divided into two classes: high-mass (HMXB) and low-mass (LMXB), each with different progenitors, evolutionary timescales, and spectral properties \citep{fabbiano2006}. A common approach to calculate XRB luminosity is via scaling relations, wherein the LMXB emission is proportional to a galaxy's stellar mass ($M_*$), and the HMXB emission is determined by the SFR \citep{Vladutescu2023, shreeram2025}. We use the relationship found by \cite{Lehmer2019A} which parameterizes the XRB luminosities as 
 \begin{equation}\label{eq:xrb_lehmer}
     L_{\text{XRB}}^{\text{gal}}=L_{\text{LMXB}}^{\text{gal}}+L_{\text{HMXB}}^{\text{gal}}=\alpha M_*+\beta SFR.
 \end{equation}
Here, the scaling parameters are based on observations of 38 nearby galaxies, setting log $\alpha$ and $\beta$ to be 29.25 (+0.07/-0.06) and 39.71 (+0.14/-0.09) respectively,  where $\alpha$ has units of erg/s/$M_\odot$ and $\beta$ has units of erg/s/$M_\odot$/yr. 

In order to simulate XRB contributions directly from our simulations, we calculate the XRB emission from each star particle directly. This methodology allows us to provide more spatial information about XRBs and robustly include contributions from satellites. We model the spectral shape as a power law, and can thus use the PowerLawSourceModel functionality in pyXSIM to generate emission from each star particle in the simulation box. We define the power law as 
\begin{equation}
N(E) = K \left( \frac{E}{E_0}\right)^{-\gamma},
\end{equation}
with $\gamma=2$ \citep{Lehmer2019A} and $E_0= 1$ keV and can derive the normalization K as follows. 
\begin{equation}
L_{\text{XRB}} = \int_{E_1}^{E_2}N(E)EdE = \int_{E_1}^{E_2}K \frac{E_0^2}{E}dE,
\end{equation}
where we integrate from $E_1 = 0.5$ keV to $E_2 = 8$ keV as that is the observed energy used for the derivation of $\alpha$ and $\beta$ in \cite{Lehmer2019A}. This can be solved for K with $L_{XRB}$ in units of keV/s:
\begin{equation}
K = \frac{L_{\text{XRB}}}{E_0^2 \text{ln}(8/0.5)},
\end{equation}
which allows us to generate maps of XRB emission directly from the simulation output. We only generate XRB stacks for the two fiducial runs, SIMBA and EAGLE, as shown in Figure \ref{fig:xrbmaps}.

Our prescription for calculating XRB emission is similar to that used in \cite{Vladutescu2023}, in that we do not model based on overall galaxy characteristics, but rather star particle by star particle. In contrast to that work, we use a slightly simpler prescription for the XRB contribution. This includes using a single power law slope for both HMXB and LMXB, as the HMXB contributions will dominate the emission. We also use the star formation rate of each particle, as opposed to their method of determining a star formation history based on assumptions about IMFs, type two supernova rates, metallicity, and stellar lifetime functions. 
We elect to use a simpler approach because the nature of our simulation outputs means we cannot capture all the intricacies of XRB behavior, especially their time-variation, and this slightly simplified model is more easily generalizable between simulations. 

We calculate the surface brightness due to XRB emission following the same methods outlined in \S \ref{xraymodel}.  

\section{Results}\label{sec4}
\subsection{CGM Properties} \label{sec4-cgm}

As the forward modeling of X-ray emission depends on the metallicity, temperature, and density of the plasma, we first explore the physical properties of hot gas in the CGM of our five simulations. Figure \ref{fig:gas_properties} shows the average mass-weighted radial profiles for galaxies with $11< \;$log$(M_*/M_\odot)<11.25$. We focus on this stellar mass bin as it contains a large enough sample so as not to be dominated by one outlier, while still containing massive galaxies that are thought to be significantly impacted by AGN feedback. 
In Appendix \ref{CGM_othermasses} we show the CGM properties of galaxies stacked by other stellar mass bins. Overall we find that the trends and conclusions we present here can be generalized to other mass ranges. 

In the top panel of Figure \ref{fig:gas_properties}, we show the average metallicity profile of our galaxies, defined as the mass fraction of any element heavier than helium normalized to a value of 0.02.  There are relatively few differences between SIMBA, EAGLE, and EAGLE-AGNdT9, but all three distribute more metals to the CGM than their respective No-AGN runs. This difference is particularly noticeable between SIMBA and SIMBA-NoAGN past a radius of 10 kpc. This can most likely be attributed to the high-velocity jets bringing material from the metal-enriched center of the galaxy, as the winds are decoupled, allowing for the material to travel large distances. The less extreme differences between the EAGLE runs with and without AGN might be due to the fact that AGN energy in EAGLE is deposited near the black hole, where it drives slower winds that have much more interaction within the galaxy where other, non-AGN related processes can have a large impact.

The central panel of Figure \ref{fig:gas_properties} shows the temperature profiles, which display more variation between simulations. As expected, the No-AGN runs show much lower temperatures past 15 kpc, as AGN activity is needed to increase thermal energy (see Appendix \ref{CGM_othermasses}, which demonstrates these differences cannot be attributed to virial heating by stacking by halo mass). Of note is the almost complete lack of difference between EAGLE and EAGLE-AGNdT9. Despite the AGNdT9 simulation depositing more thermal energy as part of its AGN activity, the impacts on CGM temperatures appear to be quite minimal. However, we do see more distinction between SIMBA and the two EAGLE runs, with SIMBA having a slightly lower temperature between 10 and 100 kpc, and a higher temperature at larger radii. This difference is most likely due to the differences in AGN feedback implementation, as SIMBA's kinetic prescription allows for a direct deposit of energy at larger radii. However, the overall difference in temperatures is quite small, never surpassing $\approx$0.1 dex in the CGM.

\nolinenumbers

\begin{table}[h!]
\centering
\begin{tabular}{|c|p{.07\textwidth}|p{.07\textwidth}|p{.07\textwidth}|}
\hline
 & \shortstack{\textbf{0-10} \\ \textbf{kpc}} & \shortstack{\textbf{75-100} \\ \textbf{kpc}} &\shortstack{ \textbf{400-500} \\ \textbf {kpc}} \\
\hline
\multicolumn{4}{|c|}{\textbf{Property: Metallicity}} \\
SIMBA: EAGLE           & 0.87 & 1.36 & 0.65
 \\
SIMBA: SIMBA-NoAGN     & 0.91 & 4.12 & 3.1 \\
EAGLE: EAGLE-NoAGN     & 0.95 & 1.28  & 2.41\\
\hline
\multicolumn{4}{|c|}{\textbf{Property: Temperature}}\\
SIMBA: EAGLE           & 1.39  & 1.02 & 1.93 \\
SIMBA: SIMBA-NoAGN     & 0.64 & 3.98 & 3.04 \\
EAGLE: EAGLE-NoAGN     &  0.82  & 2.23  & 1.44  \\
\hline
\multicolumn{4}{|c|}{\textbf{Property: Density Squared}}\\
SIMBA: EAGLE           & 3.12 & 155.45 & 429.37 \\
SIMBA: SIMBA-NoAGN     & 0.0051 & 0.050 & 0.000052 \\
EAGLE: NoAGN           & 0.03 & 0.49  & 1.96 \\
\hline
\end{tabular}
\caption{Ratio of simulation values for mass weighted metallicity, temperature, and density squared across three radial bins. } \label{tab:percdiff}
\end{table}

\linenumbers

\begin{figure}[ht!]
\includegraphics[width=0.48\textwidth]{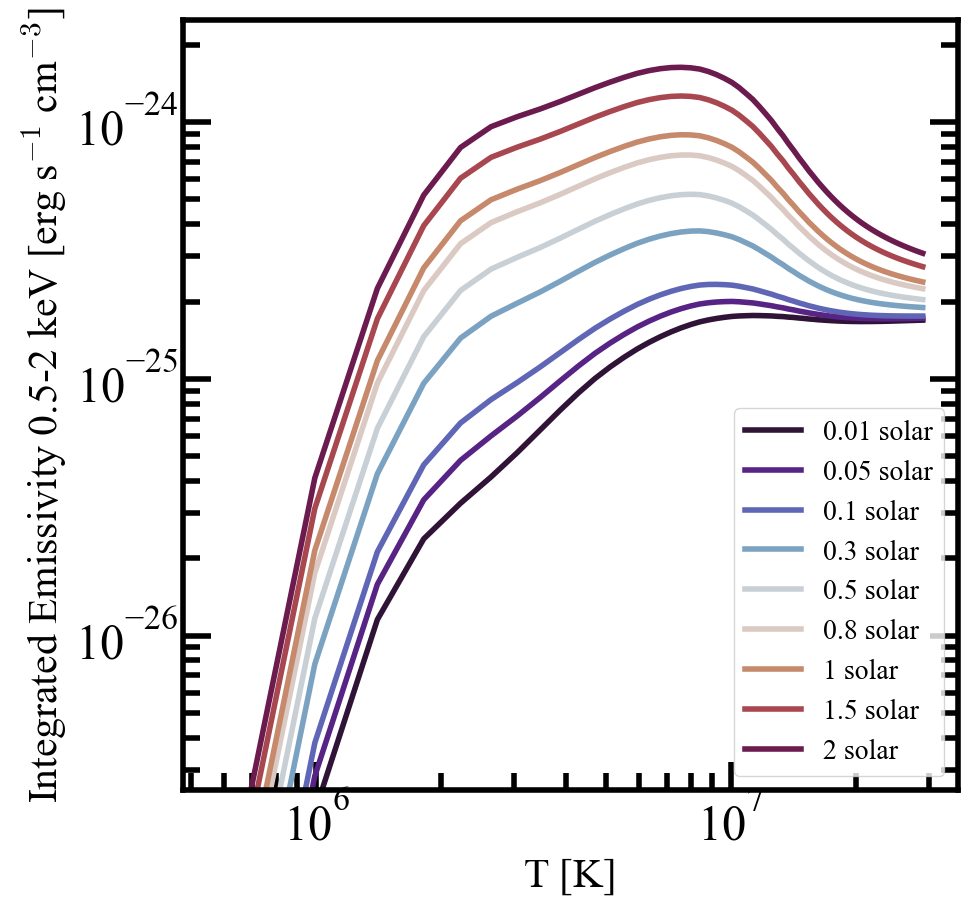}
\caption{Total soft X-ray emissivity (0.5-2.0 keV) as a function of plasma temperature and metallicity relative to solar as given by APEC and pyatomdb \citep{foster2020}. Here the different color curves represent changes in emission due to varying the O, Ne, and Fe (three species that contribute significantly to lines at this energy range) abundance relative to solar. The emission generated by pyXSIM is determined by this output from APEC and the density in the plasma.}
\label{fig:apec}
\end{figure}
\begin{figure*}[t!]
 \includegraphics[width=\textwidth]{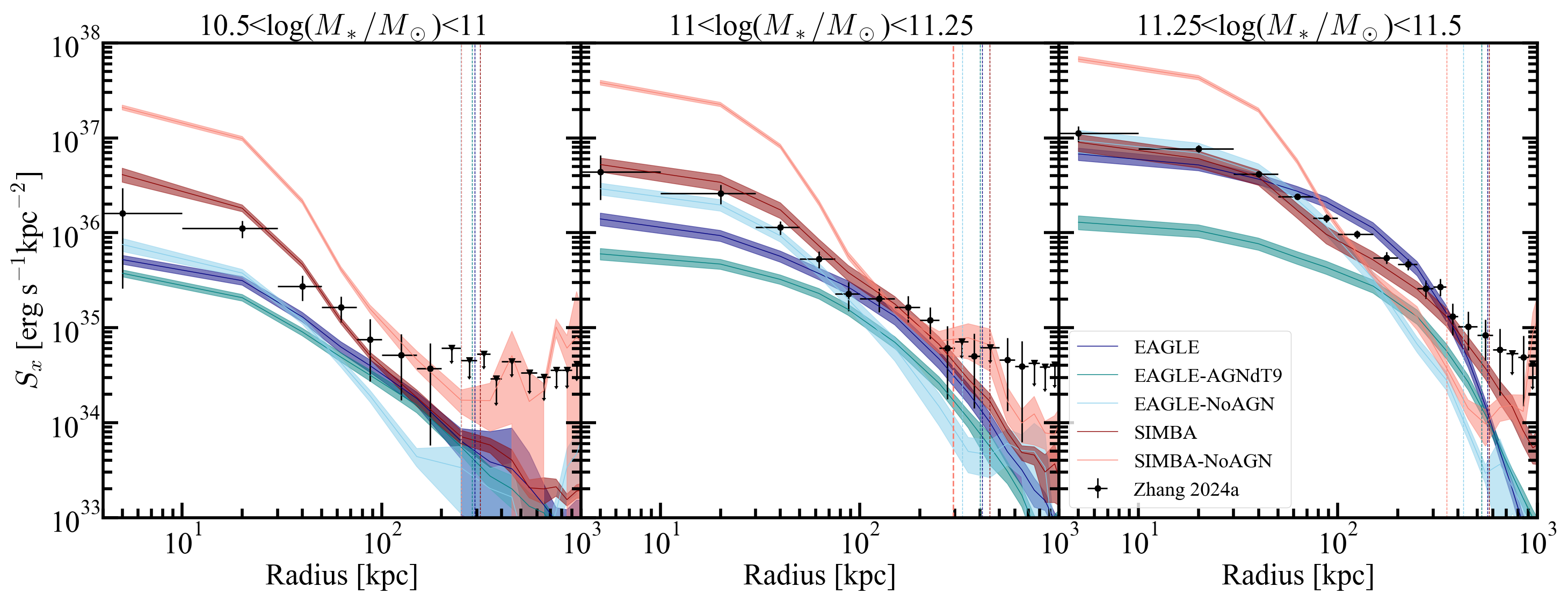}

    \caption{Soft X-ray surface brightness profiles of simulated galaxies stacked by stellar mass, compared against observed CGM emission from \cite{Zhang2024a}. Vertical dashed lines represent R$_{200c}$. From left to right, the bins are 10$<$ log(M$_*$/M$_\odot$)$<$11, 11$<$ log(M$_*$/M$_\odot$)$<$11.25, and 11.25$<$ log(M$_*$/M$_\odot$)$<$11.5. All samples have additional cuts on halo mass, such that 12 $<$ log(M$_{200c}/$M$_\odot$) $<$ 14.}
    \label{fig:comp_by_stellar_mass}
\end{figure*}

Finally, the lower panel of Figure \ref{fig:gas_properties} shows variations in density-squared. Here, the noise at large radii in the SIMBA runs represents nearby massive galaxies-although we are selecting galaxies central in their halos, there are several within this sample with a nearby massive neighbor, perhaps being in the beginning of a merger. Thus the spikes at different radii past 100 kpc represent contributions from each galaxy with a large neighbor in the stack, similar to the effect of the locally correlated environment shown in \cite{shreeram2024}. While galaxies in the EAGLE box also have some large neighbors, they are generally further away and CGM densities are more uniform, leading to fewer large spikes. This large variation can also be attributed to the lower resolution of the SIMBA simulation, which is on the order of $10^7 M_\odot$, as opposed to EAGLE's  $10^6 M_\odot$.

Of the three properties determining the X-ray emission, density has the most extreme differences between feedback prescriptions. We see that runs without AGN have higher densities in the CGM than fiducial simulations. In comparing SIMBA to the EAGLE runs, we see elevated densities in SIMBA within 20 kpc. This is most likely highly impacted by AGN feedback, as the decoupled jets in SIMBA are expected to recouple to the hydrodynamics at distances of $\approx$ 10 kpc from the center of the galaxy \citep{Simba}. While SIMBA shows a slight decrease in density around 100 kpc relative to the EAGLE runs, large contributions from neighbors serve to increase the overall density at the outer CGM. Finally, we see more of a significant difference between EAGLE and EAGLE-AGNdT9 compared to the temperature or metallicity profiles, with EAGLE showing slightly lower densities than its stronger-feedback counterpart.

Table \ref{tab:percdiff} shows the ratio in the properties explored here between different pairs of simulations in three radial bins (where the values compared are taken to be the average value in each bin). When comparing SIMBA and EAGLE, the largest differences are in the gas densities across all radial bins. The differences between fiducial and No-AGN  metallicity and temperature are larger in the outer radii bins, a trend that is not as clear in the density values. 

Figure  \ref{fig:apec} shows how soft X-ray emission is dependent on temperature and metallicity within the APEC model\citep{Smith2001, foster2020}. Overall, we see that particularly below $10^{6.2}$ K there is a strong dependence on temperature, and metallicity plays a large role at T$<10^7$ K. However, our surface brightness X-ray model is also dependent on the density squared, and given the multiple order of magnitude differences between simulations shown in the density profiles, we can expect the density to still play a very significant role in changing X-ray surface brightness in the CGM between simulations.

\subsection{Observational Comparisons}\label{sec4_1}

Figure \ref{fig:comp_by_stellar_mass} shows radial surface brightness profiles, binned into three stellar mass ranges, and compared with eROSITA observations from \cite{Zhang2024a}. Shaded regions represent the bootstrapped errors for the non-resampled set of galaxies. Comparing simulation outputs, several trends emerge. The runs without AGN feedback generally have higher surface brightness profiles in inner regions, with SIMBA-NoAGN predicting almost an order of magnitude more emission than SIMBA in all mass bins. 

For the fiducial runs, when log($M_*/M_\odot$) $<$ 11.25, EAGLE predicts a much lower surface brightness than SIMBA in the inner CGM (within $\sim 0.2 R_{200c}$ or $\sim 100$ kpc). In the highest mass bin, EAGLE and SIMBA are well aligned in all but the outermost radial bins, where EAGLE shows a steeper drop-off in surface brightness. At high masses, SIMBA and EAGLE predict very similar radial profiles, suggesting the very different AGN prescriptions have little impact on surface brightness around the highest mass galaxies, and in this regime gravitational heating is becoming a more dominant factor. However, it is in this mass bin where we see the most significant differences between EAGLE and EAGLE-AGNdT9. While EAGLE generally predicts brighter X-ray emission in the inner CGM than EAGLE-AGNdT9, the differences increase with mass, in direct opposition to the trend seen between SIMBA and EAGLE.    

We can understand these differences by returning to the CGM properties described in \S\ref{sec4-cgm}. In general, the inner CGM follows the same distribution as the density profiles. However, at larger radii the $S_X$ of the No-AGN runs drop much more than their density profiles. This can be attributed to the quick drop in temperatures seen in the middle panel of Figure \ref{fig:gas_properties}. It is worth noting that Figure \ref{fig:gas_properties} shows gas properties averaged in three-dimensional shells around each galaxy, but the surface brightness profiles are generated from 2D projections along all three lines of sight with much larger radial bins, causing the $S_X$ profiles to be more smooth. 

At log($M_*/M_\odot$) $>$ 11.5, these patterns change. As explored in Appendix \ref{CGM_othermasses} in Figure \ref{fig:gas_properties_otherbins}, at the highest mass bin differences in density and temperature between SIMBA and EAGLE are much less pronounced, accounting for the similar $S_X$ profiles. However, EAGLE-AGNdT9 does show large differences in the temperature for galaxies with 11.25 $<$ log($M_*/M_\odot$) $<$ 11.5. We could be witnessing the effects of the more bursty feedback model, as EAGLE-AGNdT9 injects more thermal energy in a given feedback event, but these events happen less frequently. If we are catching several galaxies relatively soon after a feedback event, that could account for the large spikes in temperature in the inner CGM. 

In general, none of the simulations used here consistently agree well with observations across all mass ranges. At log($M_*/M_\odot$) $<$ 11, SIMBA shows some agreement within R $\approx$ 15 kpc, but underpredicts the surface brightness at high radii. It performs much better at intermediate masses, matching the observations well within the virial radius, but underpredict surface brightness beyond that. EAGLE and EAGLE-AGNdT9 generally underpredict surface brightness for galaxies with log($M_*/M_\odot$) $<$ 11.25. For galaxies with 11.25$<$ log(M$_*$/M$_\odot$)$<$11.5, both SIMBA and EAGLE match the observations fairly well within the virial radius, although SIMBA slightly underpredicts and EAGLE slightly overpredicts $S_X$ around 100 kpc. At high radii, eROSITA observations provide only upper limits, which all of the simulations are consistent with.

\subsection{Stacking by Halo Mass}\label{sec4_2}

\begin{figure*}[ht!]
        \centering
        \includegraphics[width=1\textwidth]{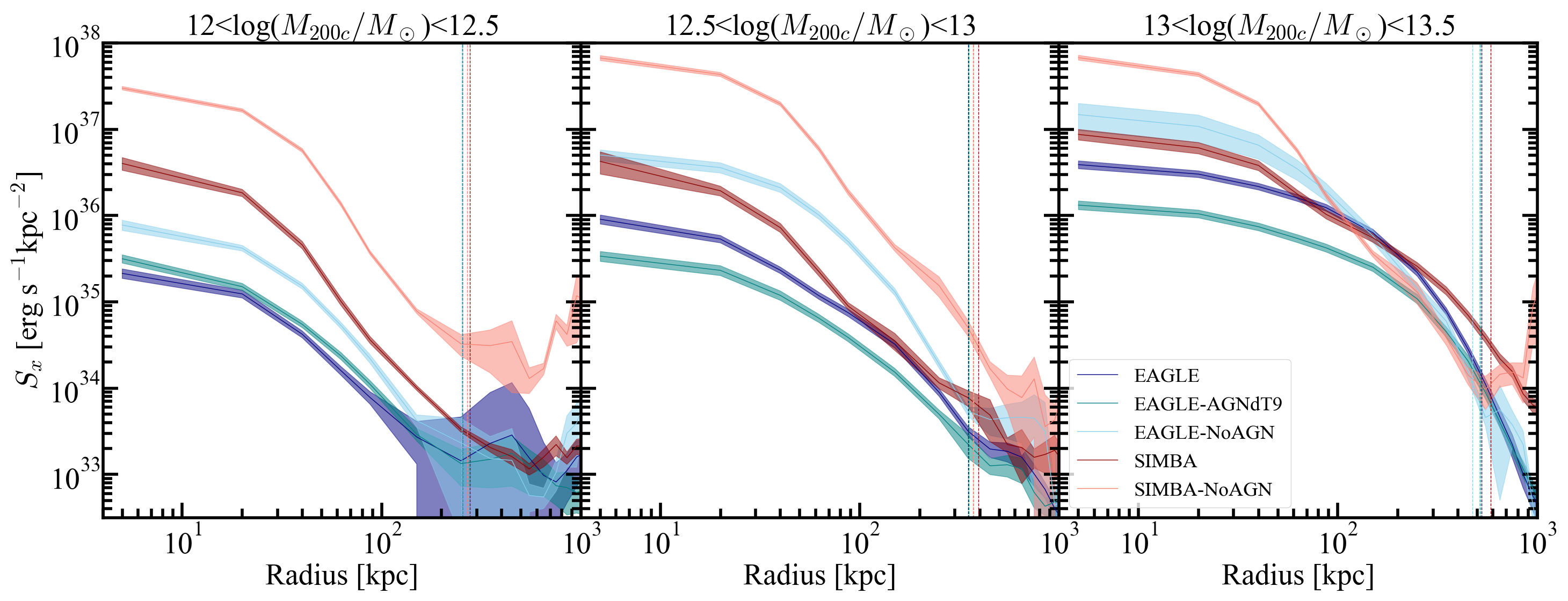}

    \caption{Radial profiles of X-ray surface brightness using galaxies from five simulations stacked by $M_{200c}$. Vertical dashed lines represent R$_{200c}$.}
    \label{fig:comp_by_halo_mass}
\end{figure*}
Figure \ref{fig:comp_by_halo_mass} shows profiles binned by halo mass. While this figure shows the same bins used in \cite{Zhang2024c}, we do not explicitly compare with observations due to the large uncertainty in stellar-halo mass relations derived observationally. Rather, our goal is to explore halo mass effects, which have been shown to be important when comparing simulations. As demonstrated in \cite{shreeram2025}, varying halo mass while maintaining the stellar mass distribution has a large impact on surface brightness profiles in Illustris-TNG. The five simulations used in this work have different stellar-halo mass relationships, particularly for the NoAGN runs, and thus halo mass bins can be quite useful in exploring differences between simulations in a more robust way. 

Here we see many of the same trends discussed in Section \ref{sec4_1}, with NoAGN runs resulting in higher surface brightness values than those with AGN, and SIMBA predicting higher $S_X$ in the inner CGM than the EAGLE runs. We also see differences between simulations decreasing at higher mass bins, similar to the stellar mass stacks. This shows that the trends that we find when comparing EAGLE and SIMBA cannot solely be attributed to differences in the halo mass distribution, but rather arise from differences in baryonic physics prescriptions and the presence of massive neighbors.

\begin{figure*}[ht!]

    \centering

        \includegraphics[width=1\textwidth]{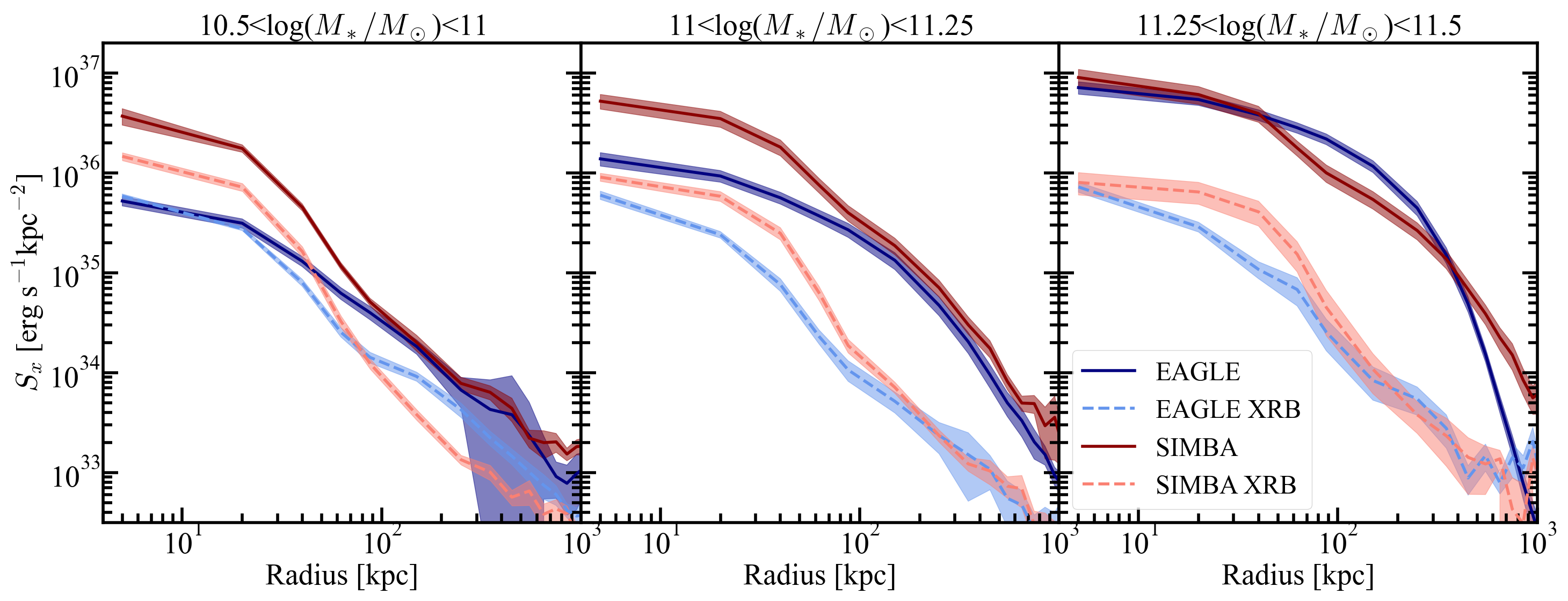}

    \caption{Surface brightness profiles of simulated galaxies stacked by stellar mass from the EAGLE and SIMBA simulations. Solid lines show emission from hot plasma, while dashed lines show forward-modelled XRB emission. From left to right, the bins are 10$<$ log(M$_*$/M$_\odot$)$<$11 , 11$<$ log(M$_*$/M$_\odot$)$<$11.25, and 11.25$<$ log(M$_*$/M$_\odot$)$<$11.5.}
    \label{fig:xrb}
\end{figure*}

\begin{table*}[ht!]
\caption{$L_x$ within $R_{500c}$ in erg s$^{-1}$ for various galaxy samples in EAGLE and SIMBA. Here we show the expected contribution from the CGM and XRBs, as well as the ratio between the two.} 
    \label{tab:LX}
    \centering
    \begin{ruledtabular}
    \begin{tabular}{cccc}

        log10($M_*/M_\odot$)  & $L_{X,\text{CGM}}$ & $L_{X,\text{XRB}}$ & $L_{X,\text{XRB}}$/$L_{X,\text{CGM}}$ \\
        \hline
         & \multicolumn{3}{c}{SIMBA} \\
        \hline
        10.5-11 & 3.5 $\pm$ 0.07 $\times$ 10$^{40}$   & 1.2 $\pm$ 0.02 $\times$ 10$^{40}$  &  0.34  \\
        11-11.25  &  1.4  $\pm$ 0.05 $\times$ 10$^{41}$  &  1.2 $\pm$ 0.04 $\times$ 10$^{40}$ & 0.09 \\
        11.25-11.5 & 4.2 $\pm$ 0.10 $\times$ 10$^{41}$ & 2.0 $\pm$ 0.05 $\times$ 10$^{40}$ & 0.05 \\
        \hline

        & \multicolumn{3}{c}{EAGLE} \\
        \hline
        10.5-11 &  1.2 $\pm$ 0.03 $\times$ 10$^{40}$   &  7.4 $\pm$ 0.07 $\times$ 10$^{39}$   & 0.59 \\
        11-11.25  & 7.8 $\pm$ 0.02 $\times$ 10$^{41}$   & 6.2$\pm$ 0.09 $\times$ 10$^{39}$ & 0.08\\
        11.25-11.5 & 6.2 $\pm$ 0.13 $\times$ 10$^{41}$  & 1.2 $\pm$ 0.05 $\times$ 10$^{40}$  &  0.02 \\

    \end{tabular}
    \end{ruledtabular}
    
\end{table*}

\subsection{XRBs}

One of the challenges when observing diffuse CGM gas in the X-ray is the presence of other X-ray sources such as AGN and X-ray binaries. As AGN emission is centered in the galaxy, its contribution to the radial surface brightness profile can be assumed to follow the PSF. XRBs are more complicated, as they can be found at larger radii and in satellites of the stacked galaxies. Different simulations can thus lead to different predicted XRB profiles, as the morphology of galaxies, their clustering, and their star formation rates are not consistent across models.

Figure \ref{fig:xrb} shows the radial profiles for CGM and XRB emission in the SIMBA and EAGLE simulations. Above log($M_*/M_\odot$) = 11, thermal emission is much stronger than XRB emission. In the lower mass bin, particularly at inner radii, XRB emission becomes more important, with EAGLE showing roughly equal surface brightness in hot plasma and XRB within 15 kpc.  When comparing the XRB emission between simulations, we see similar trends to the CGM surface brightness, with SIMBA generally predicting higher surface brightness across the two lowest mass bins but having roughly equal brightness in the highest mass bin. For both simulations, XRBs show a steeper profile than CGM, but it is worth noting that the XRB profile still contributes significant surface brightness at larger radii and the shape of the profiles when comparing EAGLE and SIMBA are different. This suggests that XRB emission cannot be treated simply as a function of the PSF, and it is not consistent between simulation models.

We can further quantify the differences between XRB and CGM emission by calculating the integrated luminosity within R$_{500c}$. This is derived from the surface brightness profile as
\begin{equation} \label{lxeq}
L_X = \Sigma^{R_{500c}}_0 S_X \times A_{\text{shell}},
\end{equation}
where $A_{\text{shell}}$ is the surface area of each radial bin. Table \ref{tab:LX} shows the average $L_{X, \text{XRB}}$ and $L_{X, \text{CGM}}$ for galaxies binned by stellar mass and star formation rate in the SIMBA and EAGLE runs. 

For both SIMBA and EAGLE, the relative contribution from XRBs is larger at lower mass bins. At 10.5 $<$ log($M_*/M_\odot$) $<$ 11, SIMBA predicts XRB luminosities $\approx 34\%$ of the CGM luminosity, with EAGLE predicting even more XRB emission, $\approx 60\%$ of the CGM luminosity. However, at the highest masses, EAGLE predicts a lower XRB/CGM ratio than SIMBA, at 2$\%$ as opposed to 5$\%$. Appendix \ref{app:xrb_profiles} provides these results for galaxies separated into star-forming and quiescent samples.

\section{Conclusions}\label{sec5}

In this work, we provide the first comparisons of EAGLE and SIMBA simulations to eRASS:4 observations. We build a model for hot circumgalactic medium emission using pyXSIM and SOXS, stack galaxies by stellar and halo mass, and present predictions for X-ray binary emission directly from star particles in the simulations. We explore how the different CGM properties in each simulation impact the predicted X-ray emission, and attempt to explain the role AGN feedback played in shaping these properties.  Our main results as follows: 

\begin{itemize}
\item We find significant differences  in surface brightness profiles between the simulations, particularly at stellar masses below $10^{11.25} M_\odot$. We show that despite the range in feedback models considered here, none of the simulations consistently aligns with observations across all radii and mass bins. For galaxies with log$(M_*/M_\odot) < $ 11.25, SIMBA outperforms EAGLE, suggesting that the decoupled kinetic wind model implemented in SIMBA is leading to more realistic CGM conditions than  purely thermal AGN feedback. At higher stellar mass, SIMBA and EAGLE predict very similar profiles, suggesting high-mass galaxy stacks are less helpful for constraining simulation physics.

\item 
We find the largest variation between simulations lies in their densities, which are very sensitive to feedback prescriptions at R$<0.2 R_{200c}$. At larger radii, neighbors/satellites contribute significantly to the gas density, making it less useful for X-ray constraints of the CGM density. However, at high radii we do see AGN feedback impacting the temperature and metallicity in the CGM. The simulations that do not include AGN feedback predict large drop-offs in the temperature and metallicity past $0.2 R_{200c}$. These drop offs are seen in the $S_X$ profiles, suggesting that X-ray surface brightness at higher radii could be a sensitive probe to the chemical and thermal enrichment of the CGM by AGN feedback.

\item Finally, we find that there are significant differences in the X-ray binary (XRB) luminosity ratios between simulations, which highlights that these simulations differ in more than just their CGM properties. At low stellar masses, these differences become especially important as EAGLE predicts roughly equal surface brightness from XRBs and CGM, especially at small radii, whereas SIMBA predicts a higher CGM $S_X$.   

\end{itemize}

Overall, these results suggest a complicated relationship between AGN feedback models, their impact on the CGM, and the resulting predicted X-ray surface brightness. The simulations used in this work are widely used, and yet fail to consistently align with observations of X-ray surface brightness in the CGM. They also suffer somewhat from low statistics, given the natural limitations that come with using a small box. However, these results do provide direction for future work that seeks to better constrain AGN feedback models with X-ray emission. The inner CGM (out to $\approx$ 0.2$R_{200c}$) provides a good place for constraining density conditions, while the outer CGM is more sensitive to temperature and metallicity. We also suggest limiting studies to stellar masses $<10^{11.25} M_\odot$, as at higher mass the differences between simulations are less significant, suggesting virial heating or other heating/environmental factors play a larger role in CGM properties than AGN feedback prescriptions. Future work exploring other CGM properties through  thermal and kinetic Sunyaev Zel'dovich measurements could also be used to help break some of the degeneracies found here.

\section*{Acknowledgements}
We would like to thank the anonymous reviewer, Seth Cohen, Darby Kramer, Phil Mauskopf, and Natalie Sanchez for their helpful discussions during the development of this work. We also thank Robert Thompson for developing CAESAR, and the yt team for the development and support of this useful community tool. S.G. acknowledges support from NSF Grant No. 2233001. E.S. acknowledges support from NASA grants 80NSSC22K1265 and 80NSSC25K7299. This research was supported by the Munich Institute for Astro-, Particle and BioPhysics (MIAPbP) which is funded by the Deutsche Forschungsgemeinschaft (DFG, German Research Foundation) under Germany's Excellence Strategy – EXC-2094 – 390783311. Support for J.A.Z. was provided by the Chandra X-ray Observatory Center, which is operated by the Smithsonian Astrophysical Observatory for and on behalf of NASA under contract NAS8-03060. MB acknowledges funding by the Deutsche Forschungsgemeinschaft (DFG) under Germany's Excellence Strategy -- EXC 2121 ``Quantum Universe" --  390833306 and the DFG Research Group ``Relativistic Jets”.

\bibliographystyle{aasjournal}
\bibliography{bib.bib}

\appendix

\section{Galaxy Sample Counts}\label{app:counts}

Tables \ref{tab:simulation_counts_Simba} and \ref{tab:simulation_counts_eagle} show the counts of galaxies by mass bin (prior to resampling) for the SIMBA and EAGLE simulation outputs, respectively. The mass bins are chosen to match those of \cite{Zhang2024c}, such that there are three stellar mass bins representing $\approx$ Milky-Way mass galaxies, M31 mass galaxies, and double M31 mass galaxies, and four halo mass bins. We find low sample sizes in the highest mass bins, particularly for runs with AGN feedback, and very low statistics once galaxies are further divided into star-forming and quiescent. 

\begin{table*}[h]
\caption{A summary of the simulated SIMBA galaxy samples, prior to resampling.} 
    \label{tab:simulation_counts_Simba}
    \centering
    \begin{ruledtabular}
    \begin{tabular}{ccccc}
        log10($M_*/M_\odot$)  & N$_{SF}$ SIMBA  & N$_{QU}$ SIMBA  & N$_{SF}$ SIMBA-NoAGN  & N$_{QU}$ SIMBA-NoAGN \\
        \hline
        10.5-11 & 297 & 375 & 387 & 6  \\
        11-11.25 & 24 & 93 &342 & 6  \\
        11.25-11.5& 6 & 36 & 189 & 0  \\
        \hline
        \hline
        log10($M_{200}/M_\odot$)  & N$_{SF}$ SIMBA  & N$_{QU}$ SIMBA  & N$_{SF}$ SIMBA-NoAGN  & N$_{QU}$ SIMBA-NoAGN\\
        \hline
        12-12.5 & 249 & 216 & 747 & 9\\
        12.5-13 & 72 & 204  & 171 & 3  \\
        13-13.5 & 6 & 84 &0 & 0 \\
        13.5-14 & 0 & 0 & 0 & 0  \\
        \hline
        
    \end{tabular}
    \end{ruledtabular}
    
\end{table*}

\nolinenumbers
\begin{table*}[h]
\caption{A summary of the simulated EAGLE galaxy samples, prior to resampling.} 
    \label{tab:simulation_counts_eagle}
    \centering
    \begin{ruledtabular}
    \begin{tabular}{p{.1\textwidth}p{.1\textwidth}p{.1\textwidth}p{.1\textwidth}p{.1\textwidth}p{.1\textwidth}p{.1\textwidth}}
        log10($M_*/M_\odot$)  &  N$_{SF}$  \newline EAGLE & N$_{QU}$  \newline EAGLE& N$_{SF}$  \newline EAGLE-  \newline AGNTd9 & N$_{QU}$  \newline  EAGLE-  \newline AGNdT9& N$_{SF}$  \newline EAGLE-  \newline NoAGN & N$_{QU}$  \newline EAGLE-  \newline NoAGN\\
        \hline
        10.5-11 &  195 & 54 & 189  & 84  &267 & 0  \\
        11-11.25  & 42 & 27 &42 & 33 & 99 & 0 \\
        11.25-11.5 & 24 &6 & 15 & 24& 69 & 0 \\
        \hline
        \hline
        log10($M_{200}/M_\odot$)  &  N$_{SF}$  \newline EAGLE & N$_{QU}$  \newline EAGLE& N$_{SF}$  \newline EAGLE-  \newline AGNTd9 & N$_{QU}$  \newline  EAGLE-  \newline AGNdT9& N$_{SF}$  \newline EAGLE-  \newline NoAGN & N$_{QU}$  \newline EAGLE-  \newline NoAGN\\
        \hline
        12-12.5 & 132 & 27 & 144 & 57 & 282 & 0\\
        12.5-13  & 87 & 42 &72 & 51 & 126 & 0 \\
        13-13.5  & 39 & 18 & 30 & 30 & 27 & 0 \\
        13.5-14 & 3 & 0 & 0 & 3 &0  & 0 \\
        \hline
        
    \end{tabular}
    \end{ruledtabular}
    
\end{table*}

\linenumbers

\section{XRB Contributions: Star-Forming Versus Quiescent}\label{app:xrb_profiles}
\begin{figure*}[t!]

    \centering
    \begin{minipage}[b]{1\textwidth}
        \centering
        \includegraphics[width=1\textwidth]{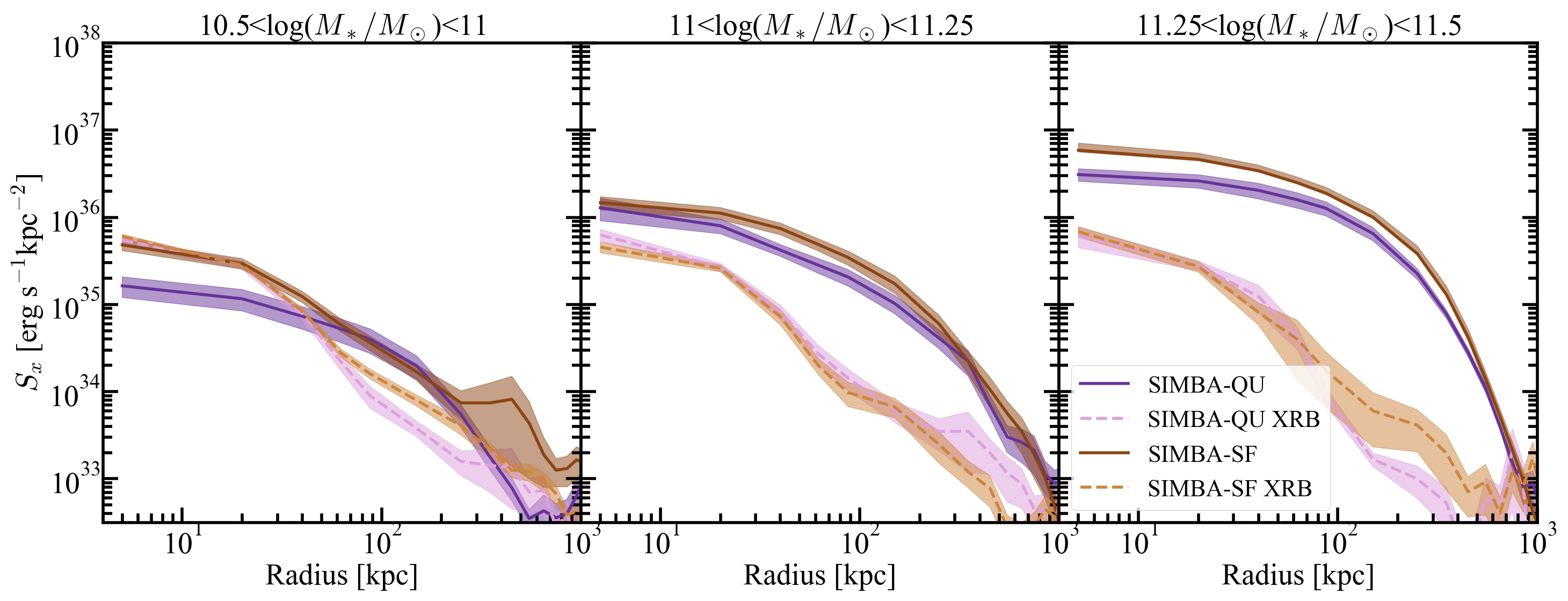}
        
    \end{minipage}
    
   \hfill
    \begin{minipage}[b]{1\textwidth}
        \centering
        \includegraphics[width=1\textwidth]{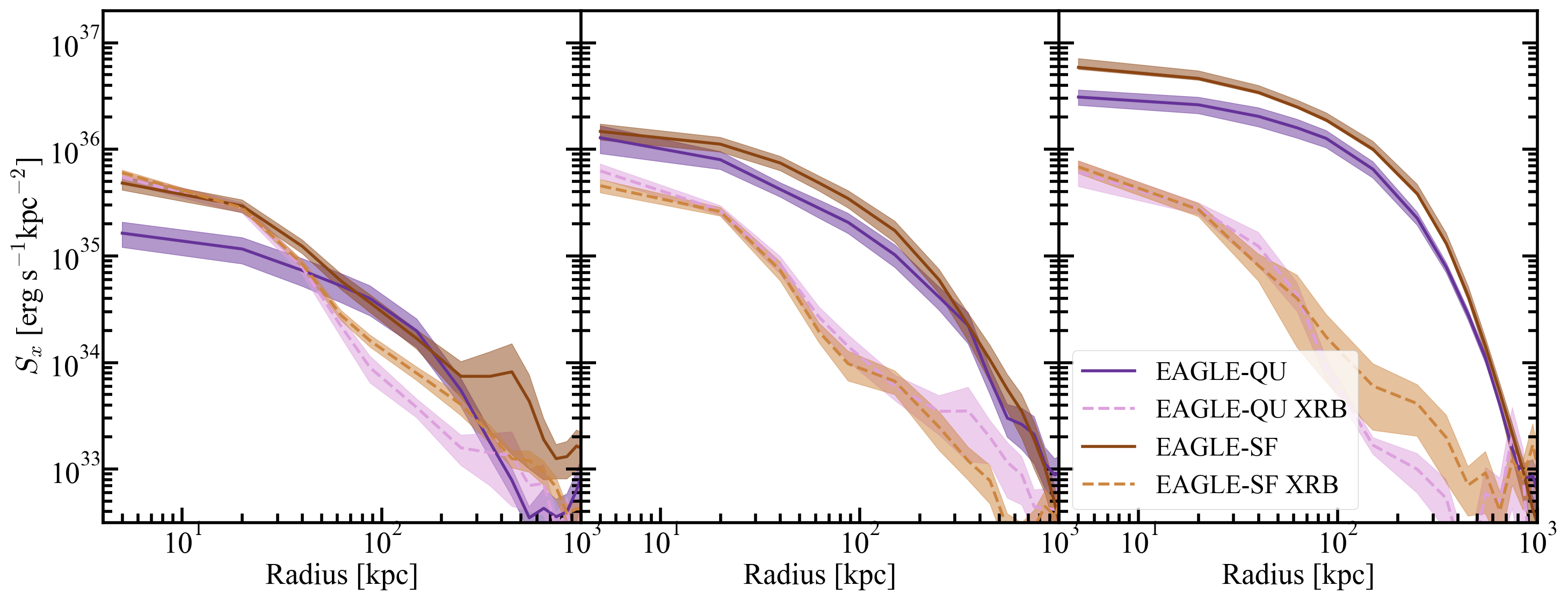}
        
    \end{minipage}

    \caption{Surface brightness profiles of simulated galaxies stacked by stellar mass and star-formation rate. Here we compare the contribution of the CGM (solid lines) to that from XRBs (dashed lines). The top row shows results from SIMBA and the bottom from EAGLE. From left to right, the bins are 10.5$<$ log(M$_*$/M$_\odot$)$<$11 , 11$<$ log(M$_*$/M$_\odot$)$<$11.25, and 11.25$<$ log(M$_*$/M$_\odot$)$<$11.5. Galaxies with a sSFR $<$ 0.01 Gyr$^{-1}$ are considered quiescent.}
    \label{fig:xrb_radial}
\end{figure*}

Figure \ref{fig:xrb_radial} shows the surface brightness profiles from XRB emission for SIMBA and EAGLE, separated by stellar mass and star formation rate, compared to the thermal emission from CGM gas. 

Table \ref{tab:LXs} shows the total luminosity within $R_{500c}$ as calculated by Equation \ref{lxeq}, separated into star-forming and quiescent galaxies. At low masses (10.5$<$ log(M$_*$/M$_\odot$)$<$11) XRB luminosities are greater for star-forming than quiescent galaxies, as expected. This trend holds for the largest mass bin (11.25$<$ log(M$_*$/M$_\odot$)$<$11.5), although with smaller differences, and in the intermediate mass bin differences between star-forming and quiescent are very small, with EAGLE predicting higher XRB luminosities for the quiescent sample. However, it is worth noting the very small sample sizes when we separate galaxies by star formation rate. Thus, we present these results as a preliminary step towards understanding how different feedback models impact star-formation rate dependent stacks. Larger box sizes will be crucial for follow-up studies that target star formation rate.

\begin{table*}[ht!]
\caption{$L_x$ within $R_{500c}$ in erg s$^{-1}$ for various galaxy samples in EAGLE and SIMBA. Here we show the expected contribution from the CGM and XRBs, as well as the ratio between the two.} 
    \label{tab:LXs}
    \centering
    \begin{ruledtabular}
    \begin{tabular}{cccc}

        log10($M_*/M_\odot$)  & $L_{X,CGM}$ & $L_{X,XRB}$ & $L_{X,XRB}$/$L_{X,CGM}$ \\
        \hline
         & \multicolumn{3}{c}{SIMBA Star-Forming} \\
        \hline
        10.5-11 & 5.3$\pm$ 0.1 $\times$ 10$^{40}$   & 1.8 $\pm$ 0.04 $\times$ 10$^{40}$  &  0.33  \\
        11-11.25  &  2.4 $\pm$ 0.17 $\times$ 10$^{41}$  &  1.9 $\pm$ 0.15 $\times$ 10$^{40}$ & 0.08 \\
        11.25-11.5 & 5.4 $\pm$ 0.10 $\times$ 10$^{41}$ & 2.3 $\pm$ 0.10 $\times$ 10$^{40}$ & 0.05 \\
        \hline
        & \multicolumn{3}{c}{SIMBA Quiescent} \\
        \hline
        10.5-11 & 1.6 $\pm$ 0.06 $\times$ 10$^{40}$  &6.1 $\pm$ 0.14 $\times$ 10$^{39}$  & 0.39  \\
        11-11.25  &  1.2 $\pm$ 0.46 $\times$ 10$^{41}$ & 1.1 $\pm$ 0.01 $\times$ 10$^{40}$ &0.09 \\
        11.25-11.5 & 4.0 $\pm$ 0.12 $\times$ 10$^{41}$ &  1.9 $\pm$ 0.05 $\times$ 10$^{40}$ & 0.05 \\
         \hline
        & \multicolumn{3}{c}{EAGLE Star-Forming} \\
        \hline
        10.5-11 &  1.3 $\pm$ 0.03 $\times$ 10$^{40}$   &  7.9 $\pm$ 0.09 $\times$ 10$^{39}$   & 0.61 \\
        11-11.25  & 9.4 $\pm$ 0.37 $\times$ 10$^{40}$   & 5.4 $\pm$ 0.11 $\times$ 10$^{39}$ & 0.06\\
        11.25-11.5 & 6.8 $\pm$ 0.16 $\times$ 10$^{41}$  & 1.3 $\pm$ 0.06 $\times$ 10$^{40}$  &  0.02 \\
         \hline
        & \multicolumn{3}{c}{EAGLE Quiescent} \\
        \hline
        10.5-11 &   1.1 $\pm$ 0.04 $\times$ 10$^{40}$ & 5.2 $\pm$ 0.09 $\times$ 10$^{39}$ & 0.5  \\
        11-11.25  & 6.4 $\pm$ 0.19 $\times$ 10$^{40}$  & 7.0 $\pm$ 0.14 $\times$ 10$^{39}$  & 0.11\\
        11.25-11.5 & 4.4 $\pm$ 0.05 $\times$ 10$^{41}$ &8.1 $\pm$ 0.10 $\times$ 10$^{39}$ & 0.02 \\
        
    \end{tabular}
    \end{ruledtabular}
    
\end{table*}

\section{Gas Properties in Other Mass Bins}\label{CGM_othermasses}
\begin{figure}[ht!]
\centering
\includegraphics[width = 0.45\textwidth]{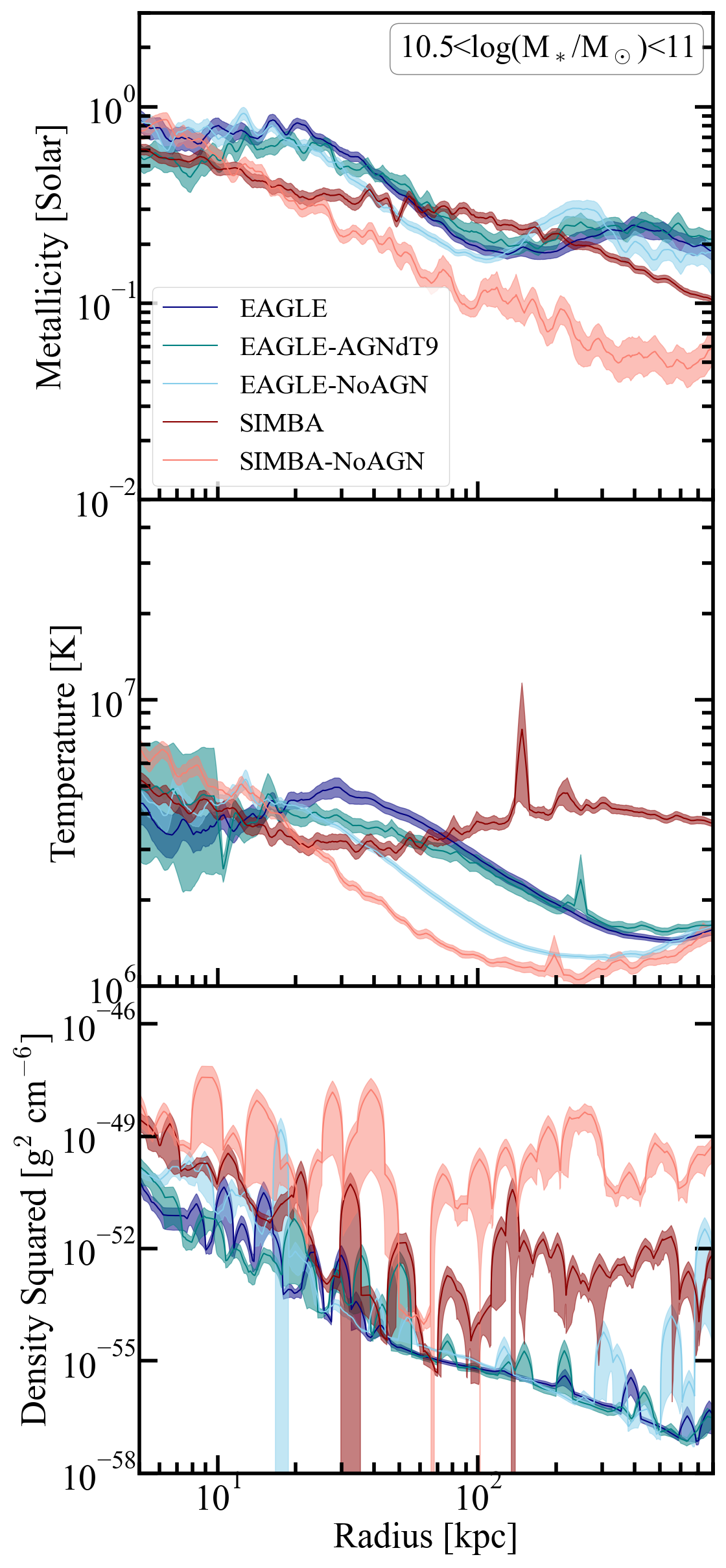}
\includegraphics[width = 0.45\textwidth]{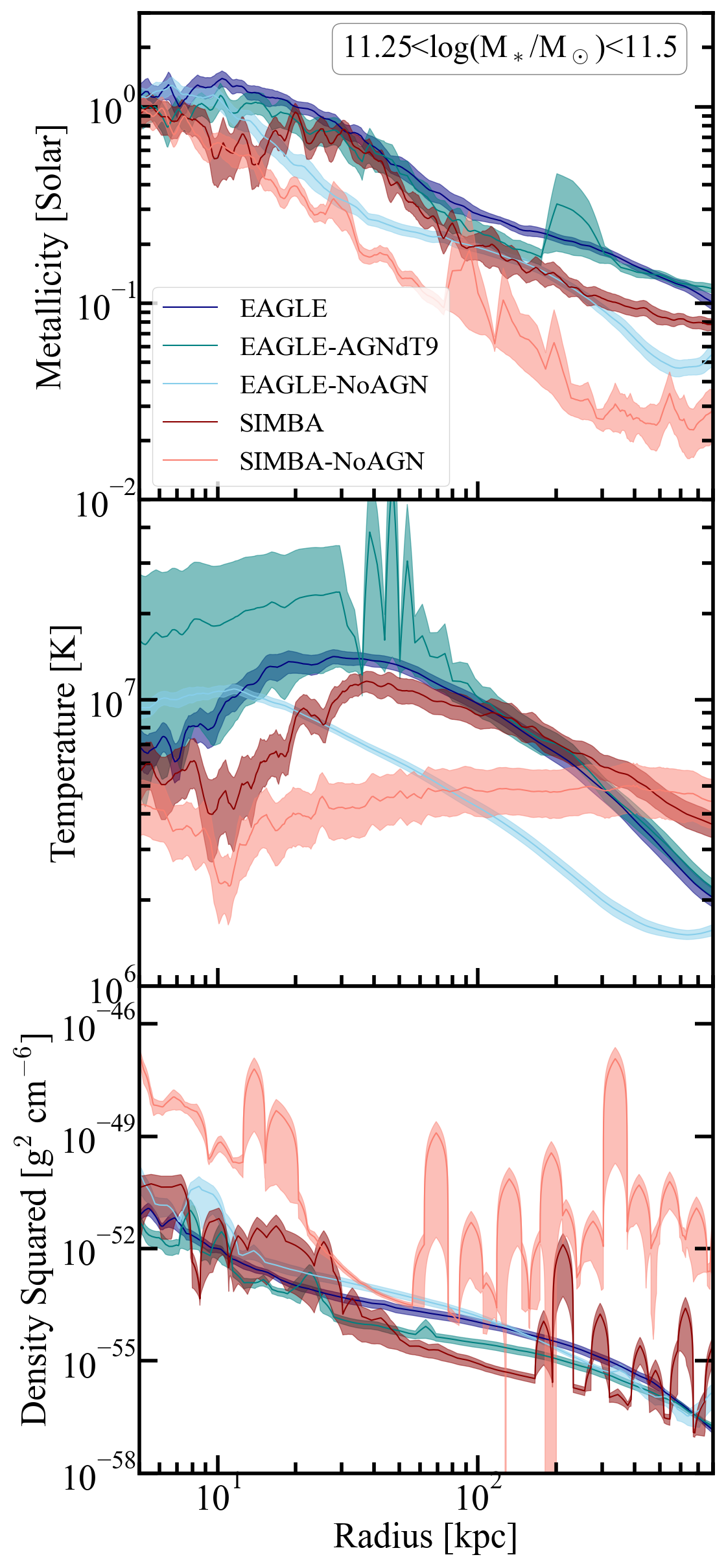}
    \caption{Radial profiles of the average mass-weighted metallicity (top), temperature (center), and density squared (bottom) of galaxies with $12<log(M_{200c}/M_\odot)<14$, and $10.5<{\rm \text{log}}(M_*/M_\odot)<11$, (left) and $12<{\rm \text{log}}(M_{200c}/M_\odot)<14$ (right). These results only include gas particles with T$>10^6$ K, as cooler particles will not significantly contribute to the soft X-ray emission (see Figure \ref{fig:apec}).}
    \label{fig:gas_properties_otherbins}
\end{figure}

Section \ref{sec4-cgm} explores the gas properties for galaxies binned by $11<log(M_*/M_\odot)<11.25$. Here we provide radial profiles of mass-weighted densities, temperatures, and metallicities for the hot plasma in the other two stellar mass bins explored in this work. Figure  \ref{fig:gas_properties_otherbins} shows galaxies stacked by $10.5<log(M_*/M_\odot)<11$ (left) and $11.25<log(M_*/M_\odot)<11.5$ (right). Overall, we see similar trends to those analyzed in Section \ref{sec4-cgm}, with the No-AGN runs having lower temperatures and metallicities, and the largest difference between simulations being found in the density profiles. 

\end{document}